\documentclass[
  twocolumn,
  prb,
  showpacs,
  amsmath,
  amssymb,
  superscriptaddress
]{revtex4}

\usepackage{bm}
\usepackage{graphicx}
\usepackage{epstopdf}

\usepackage{hyperref}
\usepackage{color}
\begin{document}
\title{Interaction induced topological protection in one-dimensional conductors}

\author{Nikolaos Kainaris}
\affiliation{Institut f\"ur Nanotechnologie, Karlsruhe Institute of Technology, 76021 Karlsruhe, Germany}
\affiliation{Institut f\"ur Theorie der Kondensierten Materie, Karlsruher Institut f\"ur Technologie, 76128 Karlsruhe, Germany}
\author{Raul A. Santos}
\affiliation{Department of Physics, Bar Ilan University, Ramat Gan 52900,
Israel }
\affiliation{Department of Condensed Matter Physics, Weizmann Institute of Science, Rehovot 76100, Israel}
\author{D.\,B. Gutman}
\affiliation{Department of Physics, Bar Ilan University, Ramat Gan 52900,
Israel }
\author{Sam T. Carr \footnote{Corresponding author\quad E-mail:~\textsf{s.t.carr@kent.ac.uk}}}
\affiliation{School of Physical Sciences, University of Kent, Canterbury CT2 7NH, United Kingdom}

\begin{abstract}
We discuss two one-dimensional model systems -- the first is a single channel quantum wire with Ising anisotropy, while the second is two coupled helical edge states.
We show that the two models are governed by the same low energy effective field theory, and interactions drive both systems to exhibit phases which are metallic, but with all single particle excitations gapped.  We show that such states may be either topological or trivial; in the former case, the system demonstrates gapless end states, and insensitivity to disorder.
\end{abstract}
\maketitle

\section{Introduction}

The importance of topology in different aspects of condensed matter physics has long been known, from topological excitations \cite{Berezinksi_1972,Kosterlitz_Thouless_1973} through topological order \cite{Wen_1995} to topological quantum computing \cite{NayakReview_2008}.  Perhaps one of the most surprising roles of topology however was the discovery of topological insulators (TIs) about a decade ago \cite{Kane_Mele_2005a,Bernevig_2006,Koenig_2007}.  The central idea is that gapped states of non-interacting electrons can be characterised by a topological index \cite{Kane_Mele_2005b}.  The original topological index for the integer quantum Hall states, the TKNN index \cite{TKKN_1982} has now been extended to all possible dimensions and underlying symmetries of the system in question, leading to a `periodic table' of possible topological insulators \cite{Kitaev_2009,Schnyder_2008}.

While the value of the appropriate topological index is a technical way of defining a topological insulator, the defining feature is the presence of edge states -- gapless excitations that live only at the edge of the sample, while the bulk of the system remains gapped.  These edge states also have the feature that they can not be localised by disorder.  They remain conducting -- perfectly so in two-dimensional topological states -- no matter how rough or disordered the edge is so long as none of the protecting symmetries are broken.

Specialising now to two dimensional systems with time reversal symmetry, we have the quantum spin Hall (QSH) effect which has a $\mathbb{Z}_2$ topological index.  Typically, this state appears in semiconductors which have band inversion due to strong spin-orbit coupling \cite{Bernevig_2006}.  The inverted band may be viewed as a negative band gap, and the defining topological index can be reduced to the sign of the gap -- a positive gap means a normal insulator, while a negative gap is the topological state.  This leads immediately to one of the main questions discussed in this work: \textit{Can one obtain states akin to a topological insulator when the gap is not present in the band structure, but dynamically generated by interactions?}  It is worthwhile noting that topological superconductors \cite{Hasan_Kane_2010} are one example of such a mechanism occurring -- after a mean field decoupling, these reduce to a non-interacting Hamiltonian and a classification as before \cite{Schnyder_2008}.  
We will discuss other examples which have no local symmetry breaking and no local order parameter, and so they cannot be so obviously put into the existing framework.

The role of interactions in topological insulators is currently a vibrant field .  One of the questions often asked is how interactions affect the metallic states at the edge of the topological insulator -- it is well understood now \cite{Xu_2006,Wu_2006,Lezmy_2012,Trauzettel_2012,Schmidt_2012,Kainaris_2014} that interactions destroy the perfect conduction predicted by the non-interacting theory due to inelastic scattering processes.  For weak to moderate interactions, this gives a temperature dependent reduction to the conductance, the perfect conductance still being realised at zero temperature.  For strong interactions however, the topological protection may be broken completely, and the edge state may localise.  The second question we will be concentrating on in this work is the opposite of this: \textit{Can interactions enhance the topological protection of the state, i.e. make it less sensitive to impurities?}

We will discuss two models that have recently been proposed that show emergent topological properties as discussed above.  The first is a model of a single (spinful) channel quantum wire with Ising anisotropy \cite{Kainaris_Carr_2015,Keselman_2015}.  The second is two coupled quantum spin Hall helical edge states \cite{Santos_2015,Santos_2016}.  In this case, it is important to note that the $\mathbb{Z}_2$ topological classification of the quantum spin Hall effect means that an even number of coupled edge states (i.e. two) are not topologically protected (in the absence of interactions).

The common feature of these two models is that they are each interacting one-dimensional systems with two channels -- in the former case, these two channels are the two spin projections; in the latter, the two distinct helical edge states.  The low-energy effective Hamiltonian is thus equivalent in both cases.  When the interactions are treated \cite{GNT_book,Giamarchi_book}, the channels rearrange themselves into two independent propagating collective modes -- one carrying charge, and the other carrying the remaining degree of freedom (spin in the former case, relative distribution of charge in the latter).  We will describe the conditions under which this charge mode remains gapless, but the other mode acquires a gap.  In line with our previous discussion, we will see in both cases that this gap can have two possible signs -- one of which is topological and exhibits edge states.

The system is however not an insulator, as there are gapless modes in the charge sector throughout the entire one-dimensional bulk.  The metallic charge sector does not escape the influence of topology however -- we show that due to the gap in the spin sector, the charge sector is robust (or even topologically protected) against backscattering from impurities in the same way as a single (topologically protected) helical edge in the quantum spin Hall effect.  This is in strong contrast to conventional wisdom of impurities in a one-dimensional Luttinger liquid, whereby the states are localised for all but the strongest attractive interactions \cite{Giamarchi1988}, while even a single impurity drives the conduction to zero for repulsive interactions \cite{Kane_Fisher_1992}.

The emergent topology therefore manifests itself in two ways -- firstly in the presence of edge states in the spin/relative charge sector, and secondly in the protection against backscattering of the gapless charge modes.  We will show at the end that these two properties are related, in that one implies the other.  We will also discuss how to potentially classify the emergent topological phase, which although inspired from the ideas of topological insulators cannot be manifestly written as a non-interacting theory.

\section{Quantum wire with Ising anisotropy}

\begin{figure}
      \begin{center}
      \includegraphics[width=0.37\textwidth]{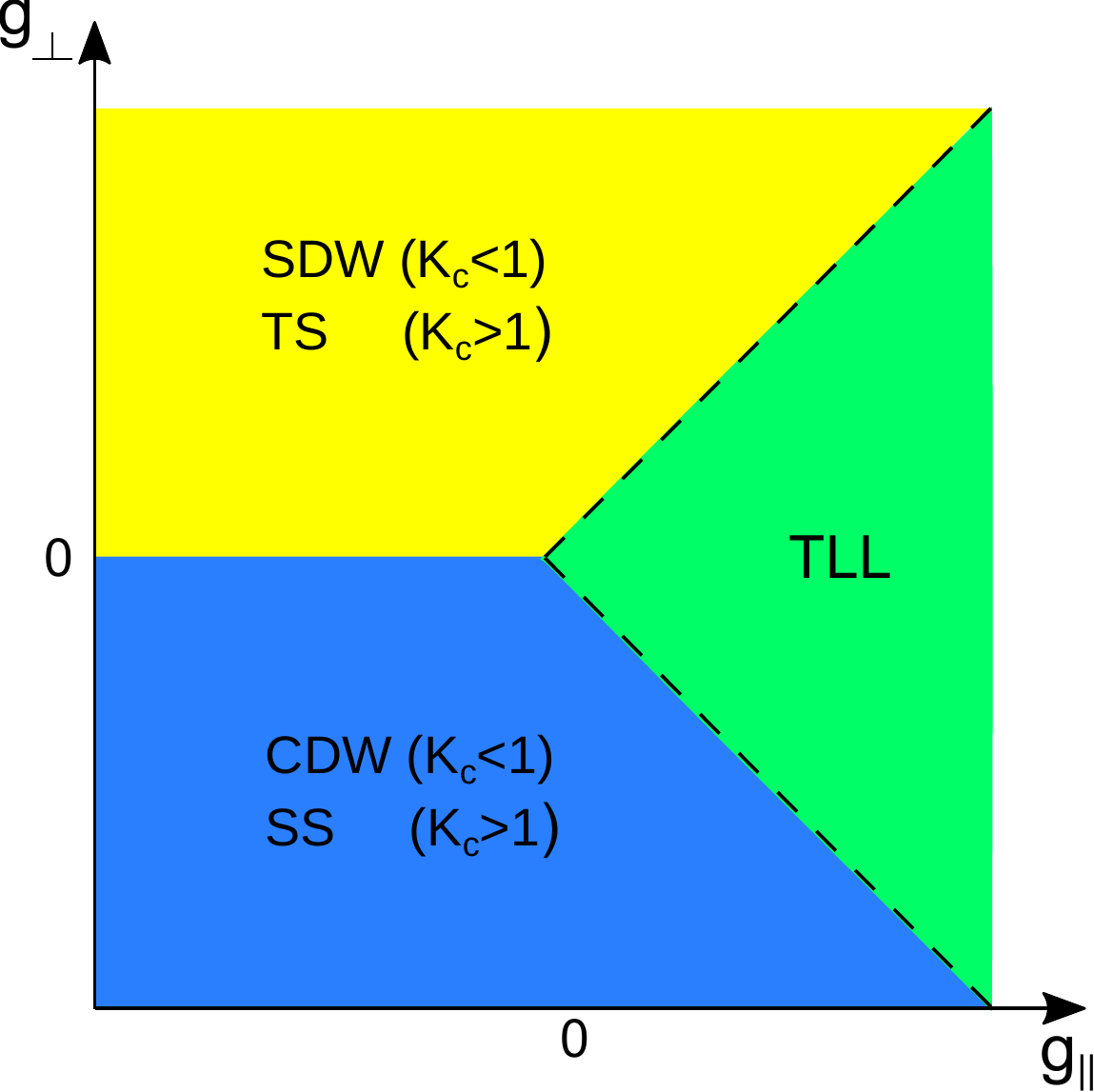}
         \caption{\small Phase diagram of a spinfull TLL. The separatrix $|g_{\perp}| = g_{\parallel}$ depicted as a dotted line belongs to the TLL phase.
          \label{Fig:phasediagram}}
      \end{center}   
\end{figure} 
\noindent
The first system we want to discuss is a single channel quantum wire. It is thereby important that the spin-rotation invariance of the electrons in the wire is broken, such that the low energy model of the system contains Ising (easy-axis) anisotropy in the spin degrees of freedom; the reason will become clear shortly.
We begin by introducing the low energy model that describes the electrons in the wire.
Thereby, it is convenient to define the vector of fermionic fields $ \boldsymbol{c}(k) = (c_{\uparrow,R} c_{\downarrow,R},c_{\uparrow,L},c_{\downarrow,L}) $, where $c_{\sigma,\eta}$(k) destroys a fermion with momentum $k$, spin $\sigma=\uparrow,\downarrow$ and chirality $\eta = R,L$. In this notation the electron Hamiltonian takes the form $H = H_0 + H_{\rm int}$, with the kinetic term
\begin{align}
   H_0 = \sum_{k} \boldsymbol{c}^{\dagger}(k) h_0(k)  \boldsymbol{c}(k) \, .
\end{align}
Here,  $h_0 = v_F k \,\sigma^0_{\sigma,\sigma'} \tau^z_{\eta,\eta'}$ is a hermitian matrix, where $v_F$ is the Fermi velocity and $\sigma^a,\tau^a$ denote the Pauli matrices in spin- and chiral space, respectively.
The interaction term is given by
\begin{align}
   H_{\rm int} = U \sum_{x} n_{\uparrow} n_{\downarrow} + V \sum_{x} \, \left(  c_{R,\uparrow}^{\dagger} c_{L,\uparrow}^{\phantom{\dagger}} c_{L,\downarrow}^{\dagger} c_{R,\downarrow}^{\phantom{\dagger}} + {\rm H.c.} \right) ,
\end{align}
with the fermionic density $n_{\sigma} = c_{\sigma,R}^{\dagger} c_{\sigma,R}^{\phantom{\dagger}} + c_{\sigma,L}^{\dagger} c_{\sigma,L}^{\phantom{\dagger}}$.
The electrons in a single channel quantum wire are a good realization of a Tomonaga Luttinger liquid (TLL). The distinctive feature of this state is that the elementary excitations are not single electrons but collective modes: charge plasmons and spinons. 
Technically, the collective nature of excitations becomes apparent under the bosonization mapping:
\begin{align}
   c_{\sigma,R} = \frac{\kappa_R}{\sqrt{2 \pi a_0}} e^{i \sqrt{4 \pi} (\varphi_{\sigma} - \theta_{\sigma})} \, , \enspace c_{\sigma,L} = \frac{\kappa_L}{\sqrt{2 \pi a_0}} e^{i \sqrt{4 \pi} (\varphi_{\sigma}+ \theta_{\sigma})} \, .
\end{align}
Here, $\kappa_{\eta}$ denote Klein factors and $a_0$ is the short distance cutoff of the field theory. The bosonic fields $\varphi_{\sigma}$ and $\theta_{\sigma}$ obey the equal time commutation relations
\begin{align}
   [\varphi_{\sigma}(x), \partial_y \theta_{\sigma}(y)] = -i \delta(x-y) \, .  
\end{align} 
Introducing the charge and spin components $\varphi_{c,s} = ( \varphi_{\uparrow} \pm \varphi_{\downarrow}) / \sqrt{2}$ the Hamiltonian density in real space decouples into charge and spin parts $\mathcal{H} = \mathcal{H}_c +\mathcal{H}_s$, where
\begin{align}
\begin{split}   
   \mathcal{H}_c =& \frac{v_c}{2} \big[ K_c (\partial_x \theta_{c})^2 + K_c^{-1}(\partial_x \varphi_{c})^2 \big]\, ,\\
   \mathcal{H}_s =& \frac{v_F}{2}  \Bigl[ (\partial_x \theta_{s})^2 + \Big( 1-\frac{g_{\parallel}}{\pi v_F}\Big) (\partial_x \varphi_{s})^2
                    \Bigr] \\
                  & +\frac{g_{\perp}}{2 (\pi a_0)^2} \cos\big( \sqrt{8 \pi} \varphi_s \big)  \, .
\label{TLLHamiltonian}                          
\end{split}                                                        
\end{align}
Here, $K_c \simeq 1 - a_0 U /2 \pi v_F$, $g_{\parallel} = a_0 U$, $g_{\perp} = a_0 V$ and $v_{c,s} = v_F/ K_{c,s}$.
The charge sector describes a Luttinger liquid with plasmon velocity $v_c$ and Luttinger parameter $K_c$. The spin sector consists of two terms, the kinetic energy, with coupling constant $g_{\parallel}$ and the potential energy with coupling constant $g_{\perp}$. The competition between these two terms determines the phase diagram of the model depicted in Fig.~\ref{Fig:phasediagram}. 
If the potential energy is large the system develops a gap in the spin sector, but is still gapless in the charge sector. The resulting strong coupling phases are thermodynamically equivalent, but have opposite signs of the spin-gap, $\Delta_s = g_{\perp}/2\pi a_0$ . 

They can be characterised by looking at potential local order parameters. As the charge mode always remains gapless, the order parameters are never nonzero in the thermodynamic limit. Rather the phase of the system is determined by the order parameter with the slowest decaying correlations.

For definiteness we consider repulsive interactions in the charge sector, $K_c <1$.
In this case we have to study two possibilities for the order parameter. For $g_{\perp} <0$ the potential energy $\sim g_{\perp} \cos(\sqrt{8 \pi} \varphi_s)$ is minimized by $\varphi_{s,n}^{\text{CDW}} \equiv \sqrt{\pi/2} n$. For these mean-field configurations, $\langle\cos(\sqrt{2 \pi } \varphi_s)\rangle \neq  0$ and the dominant correlations are of the CDW type, with the order parameter 
   \begin{align}
   \begin{split}
         \mathcal{O}_{\text{CDW}} =& \sum_{\sigma,\sigma'}  \sum_{\eta,\eta'} c_{\sigma,\eta}^{\dagger} \sigma^0_{\sigma,\sigma'} \tau^x_{\eta,\eta'}
                                     c_{\sigma',\eta'}^{\phantom{\dagger}}  \\                      
                           =& \frac{2}{\pi a_0} \sin(\sqrt{2 \pi} \varphi_{c}) \cos(\sqrt{2 \pi} \varphi_{s}) \, . 
                           \label{CDW} 
   \end{split}                        
   \end{align}  
On the other hand if  $g_{\perp} >0$ the potential energy is minimized by $\varphi_{s,n}^{\text{SDW}} = \sqrt{\pi/2} (n+1/2)$ and the order parameter
   \begin{align}
   \begin{split}
        \mathcal{O}_{\text{SDW}} =&  \sum_{\sigma,\sigma'} \sum_{\eta,\eta'}c_{\sigma,\eta}^{\dagger} \sigma^z_{\sigma,\sigma'} \tau^x_{\eta,\eta'}  c_{\sigma',\eta'}^{\phantom{\dagger}} \\
                           =& \frac{2}{\pi a_0} \cos(\sqrt{2 \pi} \varphi_{c}) \sin(\sqrt{2 \pi} \varphi_{s}) \, ,
                           \label{SDW1} 
   \end{split}                        
   \end{align}
   \begin{figure}
      \begin{center}
      \includegraphics[width=0.38\textwidth]{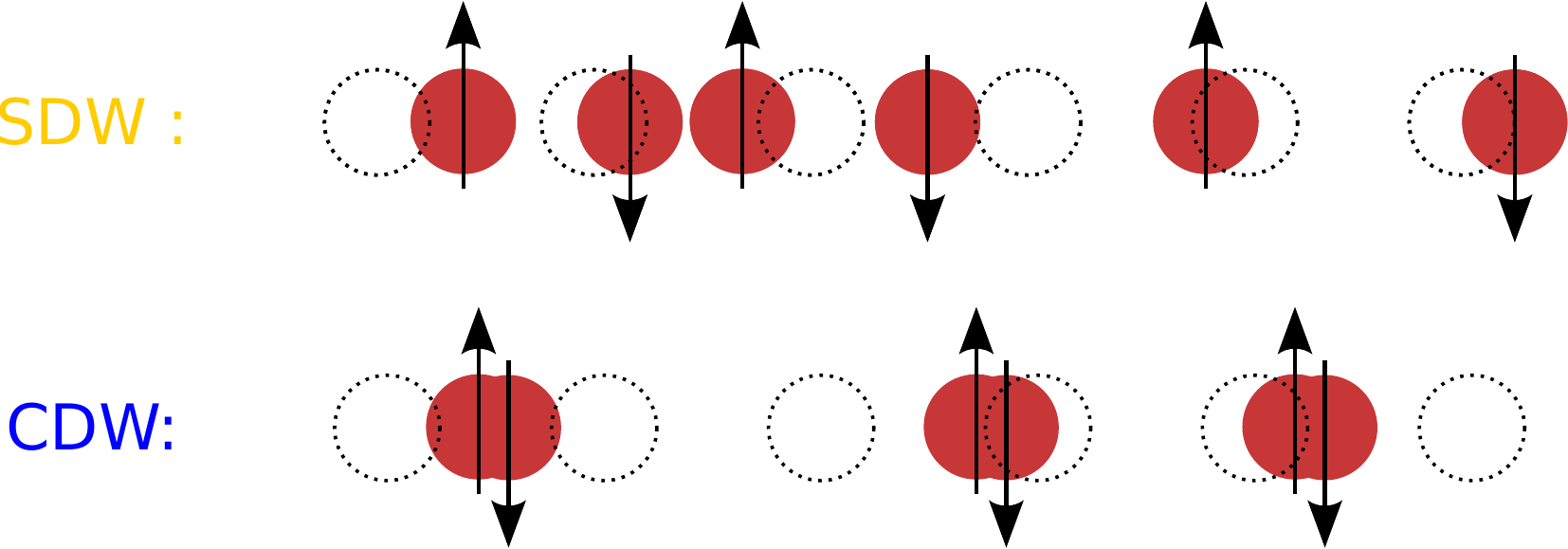}
         \caption{\small Strong coupling phases of the model in Eq.~(\ref{TLLHamiltonian}) with $K_c<1$. Spin degrees of freedom order while charge degrees of freedom (red solid circles) fluctuate around their average position.
          \label{Fig:sc_phases}}
      \end{center}   
\end{figure} 
describing the z-component of a spin density wave, becomes dominant since $\langle\sin(\sqrt{2 \pi } \varphi_s)\rangle \neq  0$. A cartoon picture of the two types of quasi-longrange order is depicted in Fig.~\ref{Fig:sc_phases}.
The analysis for attractive interactions, $K_c>1$, is analogous and in that case the dominant order parameters is triplet superconductivity (TS) for $g_{\perp}>0$ and singlet superconductivity (SS) for $g_{\perp}<0$.

We will show that one of the gapped phases (SDW or triplet SC) is topological, while the other (CDW or singlet SC) is topologically trivial. We will continue to use terminology for the $K_c<1$ phases, but since the topological properties are solely determined by the spin sector, the results hold also for the corresponding superconducting phases.

A subtle point, that is often overlooked in the literature is, that the topological SDW phase can only be realized in systems with broken spin-rotational symmetry. Technically, the SU(2) symmetry in the spin sector manifests itself by the condition that the coupling constants, $g_{\perp}=g_{\parallel} =g$, are equal and thus the system is either in the TLL phase if $g>0$ or in the CDW phase if $g<0$. Furthermore, we will show that the protection of the system against Anderson localization crucially depends on conserved time reversal symmetry (TRS), so that breaking the SU(2) symmetry by application of a magnetic field is not desired.

In the context of quantum wires with broken SU(2) symmetry in the spin sector several experimental realizations of the SDW or TS phase have been proposed.
In Ref.~\cite{Kainaris_Carr_2015} the authors considered a quasi-one-dimensional semiconducting quantum wire with strong Rashba spin-orbit coupling (SOC) and identified a parameter regime where the SDW phase forms. A possible realization of the TS phase was proposed in Ref.\cite{Keselman_2015}, where the authors studied a semiconducting quantum wire proximity coupled to a superconducting wire with SOC in a certain regime of parameters. Another realization of the TS phase may also be possible in quasi-one dimensional organic conductors where spin anisotropic interactions are believed to be present.\cite{Giamarchi1988}

Next, we discuss coupled edge states of 2D TIs as another system, where the SDW phase can emerge.

\section{Two coupled helical modes}
\label{Sec:Two helical modes}

The second system where we see the appearance of emergent topological properties are the edge states of the QSH insulator.
These edge states have a helical structure, meaning that each consists of two counterpropagating
modes with opposite spin orientation. A single helical edge mode is protected against Anderson localization by time reversal symmetry, which forbids elastic scattering between Kramers partners. On the other hand, when two sets of Kramers pairs are coupled, scattering between the states is expected to localize the edge modes.
We will show that the above situation changes drastically, when we include interaction between the edge states. 

The Hamiltonian of the system $H=H_0+H_{\rm int}$ 
then consists of the noninteracting part $H_0$ and the interaction Hamiltonian $H_{\rm int}$. Defining the vector of fermionic fields 
$\boldsymbol{c}(k)=(c_{\uparrow,1}(k),c_{\downarrow,1}(k),c_{\uparrow,2}(k),c_{\downarrow,2}(k))^T$,
where $c_{\sigma,a}(k)$ destroys a fermion in the helical mode ($a=1,2$) with a spin ($\sigma=\uparrow,\downarrow$)
and momentum $k$, the non-interacting part can be written as
\begin{equation}\label{h0}
 H_0=\sum_{k}\boldsymbol{c}^\dagger(k)h_0(k)\boldsymbol{c}(k),
\end{equation}
with the hermitian matrix 
\begin{eqnarray}\label{matrix_non_int}
h_0=\delta_{aa'}(v_F\sigma^z_{\sigma\sigma'}+\alpha_{SO}\sigma^x_{\sigma\sigma'})k-t_\perp\tau^x_{aa'}\delta_{\sigma\sigma'}.
\end{eqnarray}
Here, $\sigma^{x,y,z},\tau^{x,y,z}$ are the Pauli matrices in spin and mode space, respectively.
The Hamiltonian $H_0$ accounts for the kinetic energy (with dispersion $\epsilon_{\uparrow/\downarrow}(k)=\pm 	v_F k$), spin orbit coupling ($\alpha_{SO}$), 
and tunneling ($t_\perp$) between the helical modes.

The interaction Hamiltonian is given by
\begin{eqnarray}\label{interaction_ham}
 H_{\rm int}=U_0\sum_{x,a}n_a(x)n_a(x)+2U\sum_{x}n_1(x)n_2(x).
\end{eqnarray}
The interaction constants within the same mode and between different modes are $U_0$ and $U$ respectively. 
Under generic conditions these two constants are different  ($U_0\neq U$).
The fermion densities are defined similar to before as
$n_a(x)=c^\dagger_{\uparrow,a}(x)c_{\uparrow,a}(x)+c^\dagger_{\downarrow,a}(x)c_{\downarrow,a}(x)$.

Going to a diagonal basis and bosonizing the system \cite{Santos_2016}, the full Hamiltonian density 
$\mathcal{H}_0+\mathcal{H}_{\rm int}$ splits into two commuting parts $\mathcal{H}=\mathcal{H}_+ + \mathcal{H}_-$. The Hamiltonian density $\mathcal{H}_+$ is given by 
\begin{equation}
 \mathcal{H}_+=\frac{u_+}{2} \left[\frac{(\partial_x\varphi_+)^2}{K}+(\partial_x\theta_+)^2K\right],
\end{equation}
with Luttinger parameter $K=\sqrt{(v+g')/(v+g'+4g)}$, renormalized velocity $u_+=\sqrt{(v+g')(v+g'+4g)}$ and  $g'=a_0(U_0-U)/2\pi$.
For energies above $t_\perp$, the Hamiltonian density $\mathcal{H}_-$ is 
\begin{eqnarray}\label{Ham_spin}
 \mathcal{H}_- &=&\frac{u_-}{2}\left[(\partial_x\varphi_-)^2+(\partial_x\theta_-)^2\right]
 -\frac{g'}{\pi a_0^2} \cos(\sqrt{8\pi}\theta_-),
\end{eqnarray}
with $u_-=v-g'$.

For $g'>0$, interactions within each helical mode are stronger than between them and 
the cosine potential in (\ref{Ham_spin}) has a minimum at $\theta_-=\sqrt{\pi/2}n$ where $n\in\mathbb{Z}$.
 In this case, the order parameter 
 \begin{eqnarray}\nonumber\label{OPSN}
  \mathcal{O}_{I}&=&i\sum_{\sigma\sigma' aa'}\tilde{c}^\dagger_{\sigma a}(\tau^y)_{aa'}[\cos 2\bar{k}_F x\;\sigma^z-\sin 2\bar{k}_F x\;\sigma^y]_{\sigma\sigma'}\tilde{c}_{\sigma' a'}\\
     &=&\frac{2}{\pi a_0}\cos(\sqrt{2\pi}\theta_-)\cos(\sqrt{2\pi}\varphi_+), \label{spinnematic}
    \end{eqnarray}
 becomes dominant as $\langle\cos\sqrt{2\pi}\theta_-\rangle\neq0$. Here, $\bar{k}_F= (k_F^1+k_F^2)/2 \equiv \epsilon_F/v$ and we introduced the fermion operators $\tilde{c}_{\sigma,a}=\sum_{\sigma'}(e^{i(\beta-\pi/4)\sigma^y})_{\sigma\sigma'}c_{\sigma',a}$, with the rotation angle $2 \beta={\rm tan}^{-1}\left(\alpha_{SO}/v_F\right)$.
 
\begin{figure}[t]
	\centering
		\includegraphics[width=0.4\textwidth]{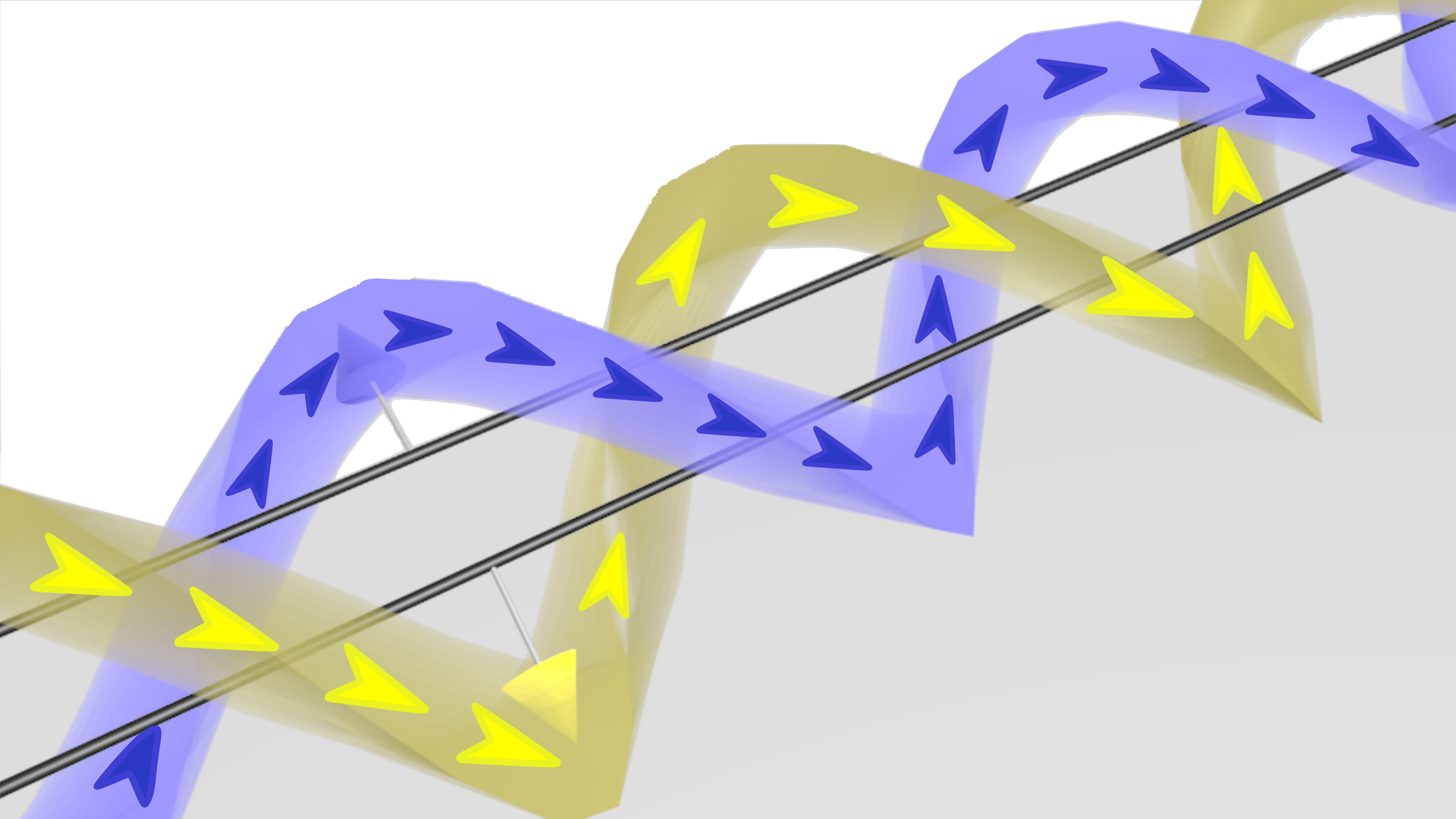}
	\caption{(color online) Patern of spin currents in the spin nematic ($g'>0$) phase, i.e. for inter-helical modes interaction
	weaker than intra mode interaction. Each spin projection guide traces a spiral along the edge, while
	also moving between the two helical modes (represented by black guides). However within these guides, the order parameter is not the spin itself, but the spin current, represented by the arrows.}\label{fig:spin_nematic}
\end{figure}

The presence of $\sigma^x$ and $\sigma^y$ (rather than $\sigma^0$) in this order parameter means that these are spin currents.
The matrix $\tau^y$ in the mode space indicates that a spin current flows between the two spin edges. The spatially dependent 
part in brackets describes a spiral for the axis of quantization of these currents. We therefore interpret this order as a 
spin-nematic phase \cite{Nersesyan1991} in the spirally varying tilted spin basis, depicted pictorially in Fig. \ref{fig:spin_nematic}

For $g'<0$ the minimum of $-g'\cos(\sqrt{8\pi}\theta_-)$ occurs at $\theta_-=\sqrt{\pi/2}\left(n+1/2\right)$ 
with $n$ integer. In this case, the order parameter 
 \begin{eqnarray}\nonumber
     \mathcal{O}_{II}&=&\sum_{\sigma\sigma' aa'}\tilde{c}^\dagger_{\sigma a}(\tau^z)_{aa'}[\cos 2\bar{k}_F x\,\sigma_z-\sin 2\bar{k}_F x\,\sigma^y]_{\sigma\sigma'}\tilde{c}_{\sigma',a'}
\\
     &=&\frac{2}{\pi a_0}\sin(\sqrt{2\pi}\theta_-)\cos(\sqrt{2\pi}\varphi_+), \label{SDW2}
    \end{eqnarray}
is dominant since $\langle\sin\sqrt{2\pi}\theta_-\rangle\neq0$. 

\begin{figure}[t]
	\centering
		\includegraphics[width=0.4\textwidth]{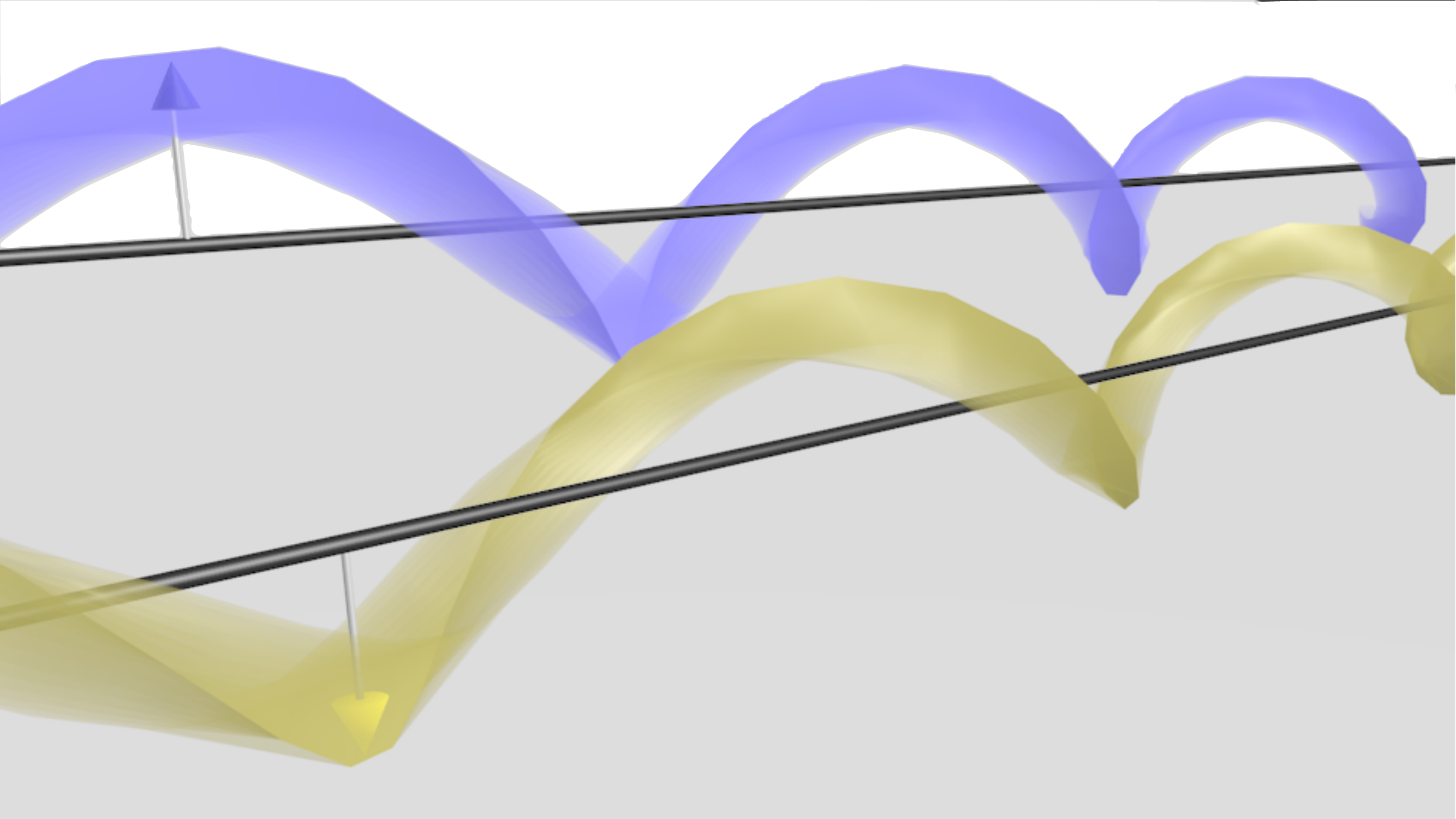}
	\caption{(color online) A pattern of spins in the SDW phase $g'<0$. Each spin projection
	traces a spiral but edge modes do not mix.}\label{fig:spin_DW}
\end{figure}

The only and main difference between this order parameter and that in Eq.\eqref{OPSN} is the presence of $\tau^z$ instead of $\tau^y$.  
This implies that one has a pattern of spins instead of spin currents, with the two different helical edges 
antiferromagnetically connected. This order parameter corresponds to a spin density wave, where as before the 
axis of quantization traces a spiral pattern along the edge of the sample. We illustrate this order parameter in Fig. \ref{fig:spin_DW}

\begin{figure}
	\centering
		\includegraphics[width=0.4\textwidth]{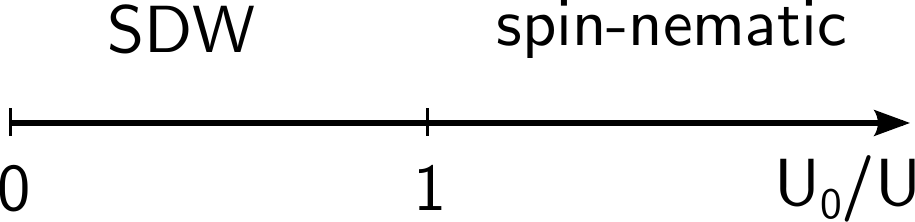}
	\caption{(color online).- Dominant order parameter for two interacting helical modes, 
	as a function of the ratio between the interaction strenghts $U_0$ (same helical mode)  vs $U$ (different helical modes)
	for strong tunneling. Above one, dominant correlations are of the spin-nematic type; below one, the dominant correlations 
	are of spin-density wave type. $H_-$ remain gapless at $U_0=U$.}\label{fig:PD}
\end{figure}

Putting these two results together, the entire phase diagram of the problem in the absence of the disorder 
is depicted in Fig.~\ref{fig:PD}.

We point out that the order parameters of the strong coupling phases in Eq.~(\ref{spinnematic}) and~(\ref{SDW2}) are completely analogous to the ones discussed in the context of quantum wires in Eq.~(\ref{CDW}) and~(\ref{SDW1}).
To connect with the results of previous sections, we include a table relating the similar operators that appear in the two physical situations.
In the following we will discuss the "topological" properties of the strong coupling phases in the context of the quantum wire setup. However, all results can be mapped to the model of helical edge modes by means of the substitutions outlined in the table.
\begin{table}[ht!]
\begin{tabular}{| l | c | c |} 
  \hline
  Model  & Quantum Wire & Edge of TI \\
\hline
  bosonic fields & $\varphi_s$ , $\varphi_c$ & $\theta_-$, $\varphi_+$ \\

\hline
  Order  & CDW ($g_\perp<0$) & Spin Nematic ($g'>0$) \\
 Parameters & SDW ($g_\perp>0$)& SDW ($g'<0$)\\
  \hline  
\end{tabular}
\caption{Comparison between the quantum wire model and the edge of a TRTI with two helical modes.}
\label{table:summary}
\end{table}

\section{Disorder}
\label{sec:disorder}

In this section we study the transport properties of the topological phases in the presence of disorder. We consider both the effect of a single impurity and random disorder and show that one strong coupling phase (CDW) is very susceptible to disorder scattering and will become localized, while the other (SDW) remains a ballistic conductor, even when disorder is added. 

We model disorder by the Hamiltonian
\begin{align}
   H_{\rm dis} = \int \! \mathrm{d} x \, \mathcal{U}(x) \boldsymbol{c}^{\dagger}(x) \boldsymbol{c}^{\phantom{\dagger}}(x) +{\rm H.c.} \, .
\end{align}
Here, $\mathcal{U}$ denotes the matrix of the disorder potential in spin and chiral space, whose entries are in general complex. For a single impurity the potential is a delta function $\mathcal{U}(x) = \mathcal{U}_{\rm imp} \delta(x)$ at the position of the impurity, say $x=0$. In the case of disorder the potential is a random matrix $\mathcal{U}_{\rm dis}(x)$, which we assume to be gaussian correlated, i.e. $\overline{\mathcal{U}^{\ast}_{dis}(x) \mathcal{U}_{dis}(y)} = \mathcal{D} \delta(x-y)$.  
The allowed matrix elements of the disorder potential are severely constricted by the symmetries of the system. First, due to time-reversal symmetry the matrix must be diagonal in spin space. Second, chiral symmetry allows for the decomposition $\mathcal{U}_{\eta,\eta'} = \mathcal{U}^{\parallel} \delta_{\eta,\eta'} +\mathcal{U}^{\perp} \delta_{\eta,\bar{\eta}}$, where $\bar{R} = L$ and vice versa. The Hamiltonian containing the forward scattering component $\mathcal{U}^{\parallel}$ can be removed by a unitary transformation.\cite{GNT_book}
The physical reason is, that forward scattering does not relax current or that impurity scattering inside the same helical edge is forbidden by time reversal symmetry, in the case of helical edges. Summarizing, we study the disorder Hamiltonian. 

\begin{align}
   H_{\rm dis} = \sum_{\sigma}\int \! \mathrm{d} x \, \mathcal{U}^{\perp}(x) c_{\sigma,R}^{\dagger}(x) c_{\sigma,L}^{\phantom{\dagger}}(x) +{\rm H.c.} \, .
\end{align}
In the bosonized form the Hamiltonian takes the form
\begin{align}
   H_{\rm dis} =- \frac{i}{4 \pi} \int \! \mathrm{d} x \, \mathcal{U}^{\perp}(x) 
                     e^{- \sqrt{2 \pi} \varphi_c} \cos(\sqrt{2 \pi} \varphi_s) +{\rm H.c.} \, . \label{disorderHamiltonian}
\end{align}
\begin{figure}
      \begin{center}
      \includegraphics[width=0.4\textwidth]{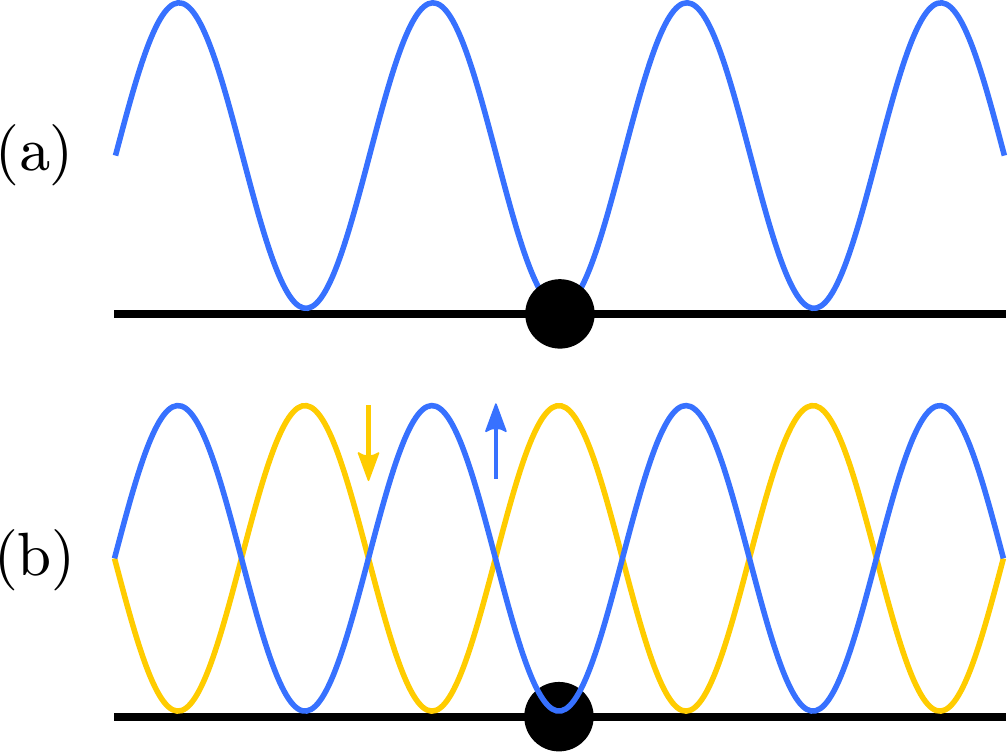}
         \caption{\small (a) Schematic showing pinning of a classical density wave by an impurity for  
         spinless electrons.  (b) In the SDW case for spinful electrons, the density waves of spin up and spin down electrons are locked out of phase.  If the impurity acts equally on the two spin projections (time reversal symmetry), then it can no longer pin the density waves.
                    \label{Fig:impurity_pinning}}
      \end{center}   
\end{figure} 
In the case of single impurity, we can directly analyze the scaling dimension of (\ref{disorderHamiltonian}). For $g_{\perp} <0$ the system is in the CDW phase and the expectation value of $\cos(\sqrt{2 \pi} \varphi_{s})$ is finite. The scaling of the impurity operator is then determined by $\sin(\sqrt{2 \pi} \varphi_{c})$ and it becomes relevant for $K_c<2$. We conclude that a single impurity in the CDW phase is a relevant perturbation that will drive the system to an insulating phase. On the other hand if $g_{\perp} >0$ the system is in the SDW phase, where the expectation value of $\cos(\sqrt{2 \pi} \varphi_{s})$ vanishes. Corrections to conductance then arise from higher order scattering processes that are generated from the impurity term under the renormalization group flow \cite{Carr2011,Carr2013}. The leading perturbation $\sim \mathcal{U}_{\rm imp}^2 \cos(\sqrt{8 \pi} \varphi_c)$, which describes coherent scattering of two electrons with opposite spin off the impurity, becomes relevant at $K_c<1/2$. Consequently, the SDW phase can be regarded as a ballistic conductor, even in the presence of an impurity, as long as interactions are not too strong. We emphasize that time reversal symmetry is crucial for the above analysis. If TRS were broken, the impurity Hamiltonian in Eq.~(\ref{disorderHamiltonian}) would contain an additional term proportional to $\cos(\sqrt{2 \pi} \varphi_c) \sin(\sqrt{2 \pi} \varphi_s)$  which becomes relevant already for $K_c<2$ and renders the SDW phase insulating.

To analyze the effect of random disorder, we average over the randomness. This yields the replicated action
\begin{eqnarray}\nonumber
 S_{\rm dis}^{AV}&=&\frac{\mathcal{D}}{(\pi a_0)^2}\sum_{\alpha\beta}\int dx d\tau_1d\tau_2\cos(\sqrt{2\pi}\varphi_s^\alpha)\cos(\sqrt{2\pi}\varphi_s^\beta)\\
 &\times&\cos(\sqrt{2\pi}[\varphi_c^\alpha-\varphi_c^\beta]).
\end{eqnarray}

Deep in the gapped SDW phase we can expand $\cos(\sqrt{2\pi}\varphi_s)$ around its minimum $\varphi_s=\varphi_{s,n}^{\rm SDW}+\delta\varphi_s$. 
Integrating out the massive $\delta\varphi_s$ mode, the model for the charge field $\varphi_c$ maps to a Giamarchi-Schultz 
\cite{Giamarchi1988} model with Luttinger parameter $K^{\rm GS}=2K_c$. Therefore the random disorder is a relevant
perturbation for $K_c<3/4$. 

Lastly, we present a semiclassical argument for the protection of the SDW phase against impurity scattering. In a spinless Luttinger liquid the excitations are charge density waves that are pinned, such that the electron density at the position of the impurity is minimized as depicted in Fig.\ref{Fig:impurity_pinning} (a). In the SDW phase the relative displacement between density waves of opposite spins, $\sqrt{2 \pi} \varphi_s$, is pinned to $\pi$. The total charge density is therefore uniformally distributed and can not be pinned by the impurity, as depicted in
Fig.\ref{Fig:impurity_pinning} (b).

\section{Topological properties}
\label{Sec:Topological properties}
\begin{figure}
      \begin{center}
      \includegraphics[width=0.48\textwidth]{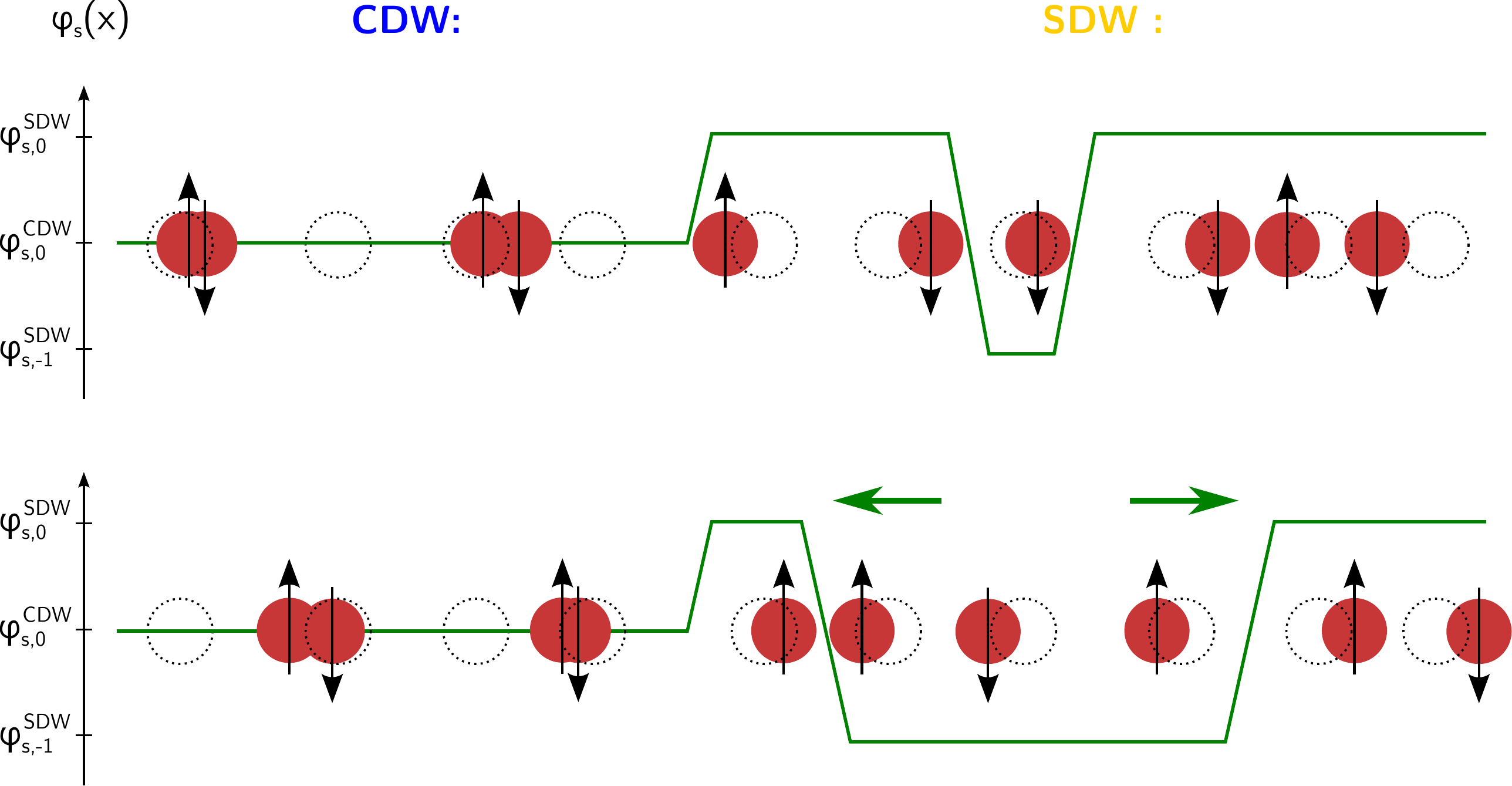}
         \caption{\small Phase boundary between topological SDW and trivial CDW phase together with a soliton-antisoliton excitation in the bulk of the SDW phase.  The green line denotes the mean field value of the spin field $\varphi_s$. 
          \label{Fig:soliton}}
      \end{center}   
\end{figure}

Besides the protection against Anderson localization in the charge sector, there is another topological property of the SDW phase: the spin sector hosts zero-energy boundary modes with fractional spin.

Let us first consider the spin sector at the exactly solvable Luther-Emery point $K_s \equiv 1+ g_{\parallel}/2 \pi v_F = 1/2$. At this special point the wave function of the edge states can be explicitly calculated by mapping the sine Gordon
model in Eq.~(\ref{TLLHamiltonian}) to a model of spinless fermions with mass
$\Delta_s = g_s / 2 \pi  a$. 
On a semi-infinite line with open boundary conditions the fermionic Hamiltonian has zero energy solutions at the boundary.\cite{Fabrizio_1995}
The wave function of the boundary mode
\begin{align}
    \chi_{0}(x) = \frac{\Delta_s}{v_s} e^{-\frac{\Delta_s}{v_s} |x|} 
\end{align}
decays exponentially into the bulk on the scale of the correlation length $\xi \sim v_s/\Delta_s$. A crucial point, that is not appreciated in the original publication, is that above solution is only normalizable if $\Delta_s >0$, i.e. the edge state only exists if the bulk is in the SDW phase.

Next, we consider a boundary between a topologically trivial phase with $\Delta_s<0$ (e.g., the vacuum) and the topologically nontrivial phase with $\Delta_s>0$ at the point $x=0$. Since the field $\varphi_s$ is pinned to $\varphi^{\text{CDW}}_{s,n_1} = \sqrt{\pi/2} n_1$ for $\Delta_s<0$ and to $\varphi^{\text{SDW}}_{s,n_2} = \sqrt{\pi/2} (n_2+1/2)$ for $Delta_s>0$ where $n_1$, $n_2$ are integers, there must be a kink of minimal magnitude $\sqrt{\pi/8}$ in $\varphi_s$ across the boundary. Such a kink in $\varphi_s$ corresponds to an accumulation of half of the electron spin at the boundary:
\begin{align}
\begin{split} 
   S_z =& \int \! \mathrm{d} x \, \rho_s(x) = \frac{1}{\sqrt{2 \pi}} \int \! \mathrm{d} x \, \partial_x \varphi_s(x) =   \pm \frac{1}{4} \, .
\end{split}       
\end{align} 
It is instructive to compare this behavior with the excitation spectrum in the bulk. The minimal excitation in the bulk is a soliton which corresponds to a transition from one minimum of the cosine potential to the next, $\varphi_{s,n}^{\text{SDW}} \to \varphi_{s,n\pm 1}^{\text{SDW}}$ with spin $S_z =\Delta \varphi_s / \sqrt{2 \pi} = \pm 1/2$. Since the edge state describes a transition from a minimum of the bulk SDW phase to the minimum of the trivial CDW phase they describe "half" of a soliton with spin $S_z = \pm 1/4$, as we found above. The soliton excitations, contrasted with a phase boundary between SDW and CDW phase, are depicted pictorially in Fig~\ref{Fig:soliton}. 

So far we have discussed the protection against disorder and the edge states of the topologically nontrivial phase separately. However, the protection against impurity scattering and the zero energy bound states at the edge are intricately connected, as we will show now.

\begin{figure}
	\centering
		\includegraphics[width=0.4\textwidth]{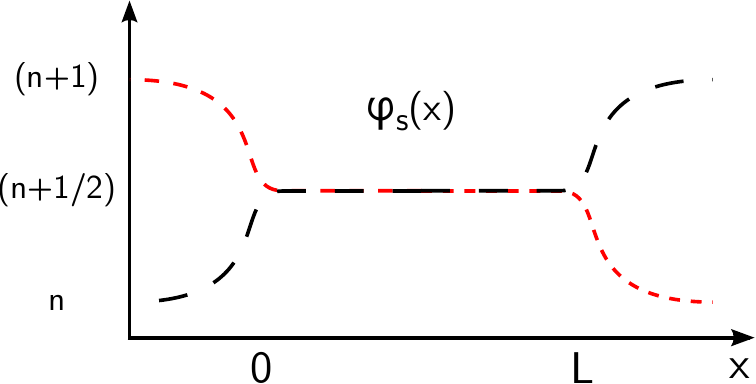}
	\caption{(color online). Spatial profile of the $\varphi_s(x)$ field in the topological phase $(g_{\perp}>0)$.
	The two different groundstates in a finite helical system correspond to the
	two choices of kink anti-kink in the boundary, where the field  has to minimize
	the backscattering potential. Different colors represent different ground state profiles for $\varphi_s(x)$.
	}\label{fig:Theta}
\end{figure}

We consider an infinite system in the topological phase with two impurities at sites $x_1=0$ and $x_2=L$. As discussed before their Hamiltonian reads as  
\begin{eqnarray}\label{potential_well}\nonumber
 U_{\rm well}&=& \sum_{\sigma,i} \int \! \mathrm{d} x \, h_w \left( c_{\sigma,R}^{\dagger}(x) c^{\phantom{\dagger}}_{\sigma,L}(x) + {\rm H.c.} \right) \delta(x-x_i) \\
&=&\left.\frac{2 h_w}{ \pi a_0} \sin(\sqrt{2 \pi} \varphi_{c}) 
                   \cos(\sqrt{2 \pi} \varphi_{s}) \right|_{x=0}^{x=L} \, .
\end{eqnarray}
The energy scale of the impurities $h_w/a_0$ is assumed to be much larger than any other scale in the problem. The potential well (\ref{potential_well})  then pins the field $\varphi_s$ to the value $\sqrt{\pi/2}m $ with $m\in\mathbb{Z}$, close to the boundary. In the bulk the field $\varphi_s$ is pinned to either 
$\varphi^{\rm CDW}_{s,n} = \sqrt{\pi/2}n$ for $g_{\perp}<0$ or 
$\varphi^{\rm SDW}_{s,n} = \sqrt{\pi/2}(n+1/2)$ for $g_{\perp}>0$. This implies that for $g_{\perp}>0$ the field $\varphi_s$ has to 
change by $\pm\sqrt{\pi/8}$ close to the boundary (see Fig. \ref{fig:Theta}). As we already discussed, this kink in the $\varphi_s(x)$ field corresponds to a spin $1/4$ excitation  near  the edge.
The two different ground states, shown in Fig.~\ref{fig:Theta}, correspond to configurations with  kink and anti-kink pairs.
Both configurations have the same energy.   This degeneracy of the $\varphi_s$ field at the edge of the samples allows particles to tunnel in or out at the edges without paying the energy cost of the gap.  These modes therefore describe the same topologically protected localized zero-mode at the boundaries of the sample that we discussed before.

\section{Discussion}
\label{Sec:Summary}
In this work we studied the emergence of topological phases in  systems that in the absence of interactions
are in a topologically trivial state. We focused on the case of repulsive  interactions, and analyzed two physical realizations:
a single channel wire with a broken $\text{SU}(2)$ symmetry and  a pair of coupled helical edge states.
Although these models may look completely different a first sight, we have shown that they are described by the same low energy effective theory.  The low energy physics is characterised by two collective modes -- one gapless, carrying charge; and the other gapped, carrying other quantum numbers which is now system specific.

In both cases, we have analyzed the possible low energy fixed points and identified  ``order parameters'', defined in this context as the
 correlation function that decays slowest with distance.  In one phase (a spin density wave in both cases), the ground state has emergent topological properties; while in the other phase, the system remains topologically trivial.  It is of crucial importance here that the gapless charge mode means that the local order parameters do not acquire a non-zero expectation value -- there is no long range order and no spontaneous symmetry breaking.  This means that there is no non-interacting mean-field description of the quasi-ordered state, which makes them rather distinct from the topological superconductors within the Bogoliubov-de-Genne universality classes \cite{Hasan_Kane_2010}.
 
Although these topological states were motivated by the non-interacting topological insulators, this last statement means that it is not obvious at all how to place them into the classification of Ref.~\cite{Schnyder_2008}.  In this framework, one looks at the properties of the Hamiltonian under certain symmetries.  While this is relatively straightforward for non-interacting topological insulators, there is an important ambiguity for the present system.  One can look at the symmetries of the original interacting Hamiltonian (which has gapless modes so is not strictly speaking an insulator), or one can exploit the spin-charge separation and study only the gapped sector of the model.  The two strategies are not equivalent: the former was applied in Ref.~\cite{Keselman_2015} putting the model in the class DIII, which is a $\mathbb{Z}_2$ topological insulator in one dimension; the latter was applied in Ref.~\cite{Kainaris_Carr_2015} putting the model in the class BDI which has a $\mathbb{Z}$ topological index.

It is easy to find potential problems with both methods.  Studying the symmetries as a whole, one can imagine, at least in principle, modifying the symmetry in a way that affects only the gapless charge sector.  This, by definition, will do nothing to the edge states in the gapped sector, and therefore should not change the symmetry classification.  On the other hand, looking only at the gapped spin sector may miss important global properties.  To begin with, the spin fields are related to the original fermionic operators in a non-local way.  This means that properties that are seemingly topological with respect to the spin field are actually local with respect to the original fermions.  More importantly, if another interaction (e.g. Umklapp terms) is added that gaps the charge sector, one is left with a fully ordered spin density wave (N\'eel) state that has long range order, spontaneously breaks time reversal symmetry, and is clearly not topological.  To the best of our knowledge, the question of topological classification of these states is unresolved at present. 

However, the physical properties resulting from this topology are unambiguous.  Edge states exist at the zero-dimensional ends of the one-dimensional systems; and the system is protected against backscattering by impurities (and hence Anderson localization) in the full one-dimensional metallic bulk.  Although the underlying physics is identical, the interpretation now depends on the system in question.  In the case of coupled helical edge states, it is well known that the topological protection may be destroyed by interactions \cite{Schmidt_2012,Kainaris_2014}.  Here, we have demonstrated the opposite -- that the lack of topological protection in two coupled non-interacting edges can be reinstated by interactions.

The case of the spin-anisotropic quantum wire is even more intriguing.  In this case, we have shown that the topological protection against impurity scattering normally associated with edge states can be achieved, even without the wire being the edge of anything.  If this state could be realised in practice, it has many potential applications in quantum electronics due to its close-to-perfect conduction properties.
 

\providecommand{\WileyBibTextsc}{}
\let\textsc\WileyBibTextsc
\providecommand{\othercit}{}
\providecommand{\jr}[1]{#1}
\providecommand{\etal}{~et~al.}


\begin{thebibliography}{[10]}

\bibitem{Berezinksi_1972}
 \textsc{V.\,L. {Berezinski{\v i}}} \jr{Soviet Journal of Experimental and
  Theoretical Physics}, \textbf{34}, 610 (1972).


\bibitem{Kosterlitz_Thouless_1973}
 \textsc{J.\,M. {Kosterlitz}} and  \textsc{D.\,J. {Thouless}}, \jr{Journal of
  Physics C Solid State Physics} \textbf{6}, 1181 (1973).


\bibitem{Wen_1995}
 \textsc{X.\,G. Wen} \jr{Advances in Physics}, \textbf{44}, 405 (1995).


\bibitem{NayakReview_2008}
 \textsc{C.~Nayak},  \textsc{S.\,H. Simon},  \textsc{A.~Stern},
  \textsc{M.~Freedman},  and  \textsc{S.~Das~Sarma}, \jr{Rev. Mod. Phys.}
  \textbf{80}, 1083 (2008).


\bibitem{Kane_Mele_2005a}
 \textsc{C.\,L. Kane} and  \textsc{E.\,J. Mele}, \jr{Phys. Rev. Lett.}
  \textbf{95}, 226801 (2005).


\bibitem{Bernevig_2006}
 \textsc{B.\,A. Bernevig},  \textsc{T.\,L. Hughes},  and  \textsc{S.\,C. Zhang},
  \jr{Science} \textbf{314}, 1757 (2006).


\bibitem{Koenig_2007}
 \textsc{M.~K\"onig},  \textsc{S.~Wiedmann},  \textsc{C.~Br\"une},
  \textsc{A.~Roth},  \textsc{H.~Buhmann},  \textsc{L.\,W. Molenkamp},
  \textsc{X.\,L. Qi},  and  \textsc{S.\,C. Zhang}, \jr{Science}
  \textbf{318}, 766 (2007).


\bibitem{Kane_Mele_2005b}
 \textsc{C.\,L. Kane} and  \textsc{E.\,J. Mele}, \jr{Phys. Rev. Lett.}
  \textbf{95}, 146802 (2005).


\bibitem{TKKN_1982}
 \textsc{D.\,J. Thouless},  \textsc{M.~Kohmoto},  \textsc{M.\,P. Nightingale},
  and  \textsc{M.~den Nijs} \jr{Phys. Rev. Lett.}, \textbf{49}, 405
  (1982).


\bibitem{Kitaev_2009}
 \textsc{A.~Kitaev}, \jr{AIP Conference Proceedings} \textbf{1134}, 22
  (2009).


\bibitem{Schnyder_2008}
 \textsc{A.\,P. Schnyder},  \textsc{S.~Ryu},  \textsc{A.~Furusaki},  and
  \textsc{A.\,W.\,W. Ludwig}, \jr{Phys. Rev. B} \textbf{78}, 195125 (2008).


\bibitem{Hasan_Kane_2010}
 \textsc{M.\,Z. Hasan} and  \textsc{C.\,L. Kane}, \jr{Rev. Mod. Phys.}
  \textbf{82}, 3045 (2010).

\bibitem{Wu_2006}
\textsc{C. Wu}, \textsc{B.\,A. Bernevig}, and \textsc{S.\,C. Zhang}, \jr{Phys. Rev. Lett.}
\textbf{96}, 106401 (2006).

\bibitem{Xu_2006}
\textsc{C. Xu}, and \textsc{J. E. Moore}, \jr{Phys. Rev. B}
\textbf{73}, 045322 (2006).

\bibitem{Lezmy_2012}
\textsc{N. Lezmy}, \textsc{Y. Oreg}, and \textsc{M. Berkooz},
 \jr{Phys. Rev. B} \textbf{85}, 235304 (2012).
  
\bibitem{Trauzettel_2012}
  \textsc{F. \,M. \,C. Cr\'{e}pin}, \textsc{J. \,C. Budich}, \textsc{F. Dolcini}, \textsc{P. Recher}, and
\textsc{B. Trauzettel}, \jr{Phys. Rev. B} \textbf{86}, 121106 (2012).

\bibitem{Schmidt_2012}
 \textsc{T.\,L. Schmidt},  \textsc{S.~Rachel},  \textsc{F.~von Oppen},  and
  \textsc{L.\,I. Glazman} \jr{Phys. Rev. Lett.} \textbf{108}, 156402
  (2012).


\bibitem{Kainaris_2014}
 \textsc{N.~Kainaris},  \textsc{I.\,V. Gornyi},  \textsc{S.\,T. Carr},  and
  \textsc{A.\,D. Mirlin} \jr{Phys. Rev. B} \textbf{90}, 075118 (2014).


\bibitem{Kainaris_Carr_2015}
 \textsc{N.~Kainaris} and  \textsc{S.\,T. Carr} \jr{Phys. Rev. B}
  \textbf{92}, 035139 (2015).


\bibitem{Keselman_2015}
 \textsc{A.~Keselman} and  \textsc{E.~Berg} \jr{Phys. Rev. B} \textbf{91},
  235309 (2015).

  
\bibitem{Santos_2015}
  \textsc{R.\,A. Santos} and \textsc{D.\,B. Gutman},
  \jr{Phys. Rev. B},
  \textbf{92}, 075135 (2015).

  
\bibitem{Santos_2016}
 \textsc{R.\,A. {Santos}},  \textsc{D.\,B. {Gutman}},  and  \textsc{S.\,T.
  {Carr}} \jr{arXiv:1601.03851} (unpublished).


\othercit
\bibitem{GNT_book}
 \textsc{A.\,O. Gogolin},  \textsc{A.\,A. Nersesyan},  and  \textsc{A.\,M.
  Tsvelik},
Bosonization and strongly correlated systems (Cambridge University Press,
  Cambridge, 1998).


\othercit
\bibitem{Giamarchi_book}
 \textsc{T.~Giamarchi},
Quantum Physics in One Dimension (Oxford University Press, Oxford, 2004).


\bibitem{Giamarchi1988}
 \textsc{T.~Giamarchi} and  \textsc{H.~Schulz}, \jr{J. Phys. France}
  \textbf{49}, 819 (1988).


\bibitem{Kane_Fisher_1992}
 \textsc{C.\,L. Kane} and  \textsc{M.\,P.\,A. Fisher}, \jr{Phys. Rev. B}
  \textbf{46}, 15233 (1992).
  
  \bibitem{Nersesyan1991}
  \textsc{A.\,A. Nersesyan}, \textsc{G. Japaridze}, and \textsc{I. Kimeridze}, \jr{J. Phys.: Condensed Matter}
  \textbf{3}, 3353 (1991).
  
  \bibitem{Carr2011}
  \textsc{S.\,T. Carr}, \textsc{B.\,N. Narozhny}, and \textsc{A.\,A. Nersesyan},
  \jr{Phys. Rev. Lett.} \textbf{106}, 126805 (2011).
  
 \bibitem{Carr2013}
  \textsc{S.\,T. Carr}, \textsc{B.\,N. Narozhny}, and \textsc{A.\,A. Nersesyan},
  \jr{Annals of Physics} \textbf{339}, 22 (2013).

\bibitem{Fabrizio_1995}
 \textsc{M.~Fabrizio} and  \textsc{A.\,O. Gogolin}, \jr{Phys. Rev. B}
  \textbf{51}, 17827 (1995).

\end{thebibliography}
\end{document}